\documentclass[journal,comsoc]{IEEEtran}

\usepackage[T1]{fontenc}% optional T1 font encoding
\usepackage{graphicx}
\usepackage{xspace}
\usepackage{cite}

\usepackage{amsmath,amssymb}
\usepackage{algorithm}% http://ctan.org/pkg/algorithms
\usepackage{algpseudocode}% loads http://ctan.org/pkg/algorithmicx
\usepackage[cmintegrals]{newtxmath}

\usepackage[utf8]{inputenc}
\usepackage{amsmath}
\newtheorem{example}{Example}
\usepackage{pgfplots}
\pgfplotsset{compat=newest}
\usepackage{tikz}
\usetikzlibrary{plotmarks}
\usetikzlibrary{external}
\usetikzlibrary{calc}
\newlength{\MyFigureWidth}
\newlength{\MyFigureHeight}
\pgfplotsset{
	every axis legend/.append style={
		legend cell align=left,
		align=left,
		font=\footnotesize
	}
}

\pgfplotsset{
	every axis plot/.append style={
  		line width=1pt,
  		mark size=1.5pt,
      mark options={solid,line width=0.5pt,fill=white!80!.}
  	}
}

\pgfplotsset{
	every axis/.append style={
        scaled ticks = false, 
        tick label style={/pgf/number format/fixed},
		label style={font=\footnotesize},
        tick label style={font=\footnotesize}  
    }
}

\definecolor{UniformColor}{rgb}{0.00000,0.44700,0.74100}%
\definecolor{CapacityColor}{rgb}{0,0,0}%
\definecolor{CCDMInfColor}{rgb}{0,0.6,0}%
\definecolor{MPDMShortColor}{rgb}{0.85,0.325,0.098}%
\definecolor{MPDMMediumColor}{rgb}{0.929,0.694,0.125}%
\definecolor{MPDMLongColor}{rgb}{1,0,0}%

\pgfplotsset{Uniform/.style={dashed,color=UniformColor}}
\pgfplotsset{Capacity/.style={solid,line width=2pt,color=CapacityColor}}
\pgfplotsset{CCDMInf/.style={dotted,color=CCDMInfColor}}
\pgfplotsset{MPDMShort/.style={solid,color=MPDMShortColor, mark=diamond*}}
\pgfplotsset{MPDMMedium/.style={solid,color=MPDMMediumColor,mark=pentagon*}}
\pgfplotsset{MPDMLong/.style={solid,color=MPDMLongColor,mark=*}}

\newcommand{\MyVec}[1]{\underline{#1}}
\newcommand{\Entr}[1]{\ensuremath{\mathbb{H}\left(#1\right)}}

\usepackage{mathtools}
\DeclarePairedDelimiter\abs{\lvert}{\rvert}%
\let\norm\relax
\DeclarePairedDelimiter\norm{\lVert}{\rVert}%
\let\floor\relax
\DeclarePairedDelimiter\floor{\lfloor}{\rfloor}
\makeatletter
\let\oldabs\abs
\def\abs{\@ifstar{\oldabs}{\oldabs*}}
\let\oldnorm\norm
\def\norm{\@ifstar{\oldnorm}{\oldnorm*}}
\let\oldfloor\floor
\def\floor{\@ifstar{\oldfloor}{\oldfloor*}}
\makeatother

\newcommand{\Npermsfun}[1]{\ensuremath{M\!\left( #1 \right)}\xspace}
\newcommand{\floortwo}[1]{\ensuremath{\floor{#1}_2}}
\newcommand{\DMfun}[1]{\ensuremath{f_{\text{ccdm}}( #1 )}\xspace}

\newcommand{\Lout}{\ensuremath{n}\xspace}
\newcommand{\Lin}{\ensuremath{k}\xspace}
\newcommand{\Nperms}{\ensuremath{N_\text{perms}}\xspace}

\newcommand{\Rateloss}{\ensuremath{R_\text{loss}}\xspace}
\newcommand{\Aquant}{\ensuremath{\tilde{A}}\xspace}
\newcommand{\TypeA}{\ensuremath{P_{\tilde{A}}}\xspace}
\newcommand{\Composition}{\ensuremath{C}\xspace}
\newcommand{\CompositionTypical}{\ensuremath{C_\text{typ}}\xspace}
\newcommand{\CompositionAcc}{\ensuremath{C_\text{acc}}\xspace}
\newcommand{\CompositionComplement}{\ensuremath{\overline{C}\xspace}}
\newcommand{\CompositionPair}{\ensuremath{\{\Composition_l, \CompositionComplement_l\}}\xspace}
\newcommand{\Npairs}{\ensuremath{N_\text{pairs}}\xspace}
\newcommand{\NpairsReduced}{\ensuremath{N^{\dagger}_\text{pairs}}\xspace}
\newcommand{\Ncompositions}{\ensuremath{N_\text{comp}}\xspace}
\newcommand{\Rshaping}{\ensuremath{R_\text{sh}}\xspace}
\newcommand{\Effshaping}{\ensuremath{\eta}\xspace} %_\text{DM}
\newcommand{\RFEC}{\ensuremath{R_\text{FEC}}\xspace}
\newcommand{\InfoRcombined}{\ensuremath{\text{I}_\text{PAS}}\xspace}

\newcommand{\AIRFiniteDM}{\ensuremath{\text{AIR}_\text{DM}}\xspace}
\newcommand{\Rbmd}{\ensuremath{\text{R}_\text{BMD}}\xspace}

\newcommand{\SetA}{\ensuremath{{\mathcal{A}}}\xspace}
\newcommand{\CardA}{\ensuremath{\abs{\mathcal{A}}}\xspace}
\newcommand{\Seqx}{\ensuremath{\MyVec{x}^n}\xspace}

\usepackage[hidelinks]{hyperref} % hyperref (almost) always goes last

\begin{document}
\title{Multiset-Partition Distribution Matching}

\author{Tobias~Fehenberger,~\IEEEmembership{Member,~IEEE},
        David S. Millar,~\IEEEmembership{Member,~IEEE}, Toshiaki Koike-Akino,~\IEEEmembership{Senior Member,~IEEE}, Keisuke Kojima,~\IEEEmembership{Senior Member,~IEEE}, and Kieran Parsons,~\IEEEmembership{Senior Member,~IEEE}
        
\thanks{T. Fehenberger was with Mitsubishi Electric Research Laboratories. He is now with Technical University of Munich, Germany. E-mail: tobias.fehenberger@tum.de.}% <-this % stops a space
\thanks{D. S. Millar, T. Koike-Akino, K. Kojima and K. Parsons are with Mitsubishi Electric Research Laboratories. E-mails: millar@merl.com; koike@merl.com; kojima@merl.com; parsons@merl.com.}}

\markboth{Fehenberger \MakeLowercase{\textit{et al.}}: Multiset-Partition Distribution Matching}%
{}

\maketitle

\begin{abstract}
Distribution matching is a fixed-length invertible mapping from a uniformly distributed bit sequence to shaped amplitudes and plays an important role in the probabilistic amplitude shaping framework. With conventional constant-composition distribution matching (CCDM), all output sequences have identical composition. In this paper, we propose multiset-partition distribution matching (MPDM) where the composition is constant over all output sequences. When considering the desired distribution as a multiset, MPDM corresponds to partitioning this multiset into equal-size subsets. We show that MPDM allows to address more output sequences and thus has lower rate loss than CCDM in all nontrivial cases. By imposing some constraints on the partitioning, a constructive MPDM algorithm is proposed which comprises two parts. A variable-length prefix of the binary data word determines the composition to be used, and the remainder of the input word is mapped with a conventional CCDM algorithm, such as arithmetic coding, according to the chosen composition. Simulations of 64-ary quadrature amplitude modulation over the additive white Gaussian noise channel demonstrate that the block-length saving of MPDM over CCDM for a fixed gap to capacity is approximately a factor of 2.5 to 5 at medium to high signal-to-noise ratios (SNRs).
\end{abstract}

\begin{IEEEkeywords}
Distribution Matching, Probabilistic Amplitude Shaping, Coded Modulation.
\end{IEEEkeywords}

\IEEEpeerreviewmaketitle

\section{Introduction}

  %% benefit of shaping
  The combination of high-order modulation, such as quadrature amplitude modulation (QAM), and strong binary codes (such as turbo-codes \cite{Berrou1993ICC_TurboCodes} or low-density parity-check codes \cite{MacKay1997ElecLett_LDPCCodes}) that operate within a fraction of a decibel (dB) of the additive white Gaussian noise (AWGN) channel capacity \cite{ChungUrbank2001CommLetters_LDPC00045dB} have become standardized in many digital communication systems. Bit-interleaved coded modulation (BICM) has achieved near universal adoption, due to its low complexity and close-to-optimal performance \cite{Caire1998TransIT_BICM}. Most coded modulation systems employ uniform signaling where each constellation point is sent with equal probability. A method to further increase the information rates is to employ constellation shaping, which gives signal-to-noise-ratio (SNR) improvements of up to 1.53~dB for the AWGN channel \cite[Sec.~4.1.3]{Fischer2005Book_Shaping} \cite[Sec.~IV-B]{Forney1984JSEL_Shaping} \cite[Sec.~IV-B]{Forney1998TransIT_ModulationCodingAWGN}. In general, there are two flavors to constellation shaping, which are geometric shaping (with equiprobable symbols drawn from an irregularly spaced constellation), and probabilistic shaping (non-uniformly distributed symbols with a regular constellation). We focus on probabilistic shaping in this paper.

  %% PAS
  Various techniques have been devised to integrate probabilistic shaping into a coded modulation system (see \cite[Sec.~II]{Boecherer2015TransComm_ProbShaping} for a review). Recently, probabilistic amplitude shaping (PAS) \cite{Boecherer2015TransComm_ProbShaping} has been proposed in which the shaping blocks are placed outside the forward error correction (FEC) encoder and decoder (see Fig.~\ref{fig:block_diagram}). This reverse-concatenation principle allows a seamless integration into existing BICM systems that, for complexity reasons, typically employ binary FEC and avoid demapper-decoder iterations. Since its proposal, PAS has attracted a lot of attention, particularly in fiber-optic communications \cite{Fehenberger2015OFC_ProbShaping,Buchali2016JLT_ProbShapingExp,Fehenberger2016JLT_ShapingQAM,Ghazisaeidi2016ECOC_ShapingPDP,Qu2018OFC_GSPSHybrid}. We focus on PAS as shaping framework in this paper.

  %% CCDM: rate loss and sequential algorithm
  An integral subsystem of a PAS system is the mapping function from the uniformly distributed data bits to shaped amplitudes and its inverse mapping. In \cite[Sec.~V]{Boecherer2015TransComm_ProbShaping}, constant-composition distribution matching (CCDM) is employed.\footnote{Note that PAS is not restricted to the use of algebraic distribution matchers (DMs) such as CCDM. %Potential alternatives are, for example, lookup tables (whose size could become prohibitively large) or shell mapping (which has limited granularity) \cite[Sec.~4.3]{Fischer2005Book_Shaping}.}
  } In simplified terms, CCDM is a fixed-length invertible operation that maps a block of Bernoulli-$\frac{1}{2}$ distributed data bits to a sequence of shaped amplitudes \cite{Schulte2016TransIT_DistributionMatcher}. Under the constant-composition principle, each output sequence must have an identical empirical distribution. The design of distribution matchers that allow to approach channel capacity is closely related to homophonic coding, see the recent review \cite{Mondelli2018TransIT_AchieveCapacity} for details.

  %% CCDM problems
  Although the principle is straightforward, designing CCDMs suitable for real-time processing is challenging, largely for the following two reasons. Any finite-length DM fundamentally suffers from a rate loss that increases with decreasing length. Hence, it would be beneficial to have CCDM block lengths of 500~shaped output symbols or more, as can be seen from, e.g., \cite[Fig.~2]{Schulte2016TransIT_DistributionMatcher}. The most widely used algorithm for implementing CCDM is arithmetic coding \cite[Sec.~IV]{Schulte2016TransIT_DistributionMatcher}, which is an inherently sequential algorithm. The combination of sequential mapping and long block lengths currently makes a real-time implementation of CCDM a highly challenging task, particularly in the context of optical fiber communications where symbol rates may be 30~GBaud or more. Hence, improved CCDM algorithms with reduced serialism must be devised, or architectures must be sought that allow to reduce the block length at equal performance. This paper is devoted to the latter.
  %In this paper, we propose a distribution matching scheme that uses arithmetic coding, yet requires significantly shorter block lengths than conventional CCDM for a given rate loss.

  %% contributions: non-cc DMs
  In this work, we examine distribution matching techniques for which the constant-composition property of conventional CCDM is lifted. Non-constant-composition DMs are based on the principle that the \emph{ensemble average} over all output sequences must have the desired composition, as opposed to the CCDM principle that every output has identical composition. While the removal of this constraint enables large gains over CCDM in the range of block lengths where brute-force computation or numerical optimization are feasible, these techniques are impossible in the block-length regime where low absolute rate loss is achievable. For example, a DM with a sequence length of merely 10~symbols and 1.5~bits of entropy for 4 shaped amplitudes will select $2^{15}$ sequences from a possible $2^{20}$. Due to the scaling of this combinatorial problem, it is clear that we need to have a constructive algorithm for efficiently generating non-constant-composition distribution matchers.
  
  %% contributions: partitions
  We propose multiset-partition distribution matching (MPDM), which forms output sequences that are equal-length partitionings of a multiset with the desired distribution.
  % This problem can be related to integer partitioning \cite{Wilf2000_LecturesonIntegerPartitions}.
  In the binary-partitioning case, for example, we consider sequences which pairwise follow the target distribution. This set of sequences will have the set of CCDM sequences as a subset, and is therefore a generalization of CCDM. We demonstrate numerically that MPDM requires significantly shorter block lengths than CCDM at a particular rate loss, which facilitates realization in hardware.

  MPDM can be viewed as a well-balanced combination of multiple CCDM instances that each have the same alphabet size as the target distribution. It is thus fundamentally different from bit-level distribution matching \cite{Boecherer2017Arxiv_PDM,Pikus2017CommLetters_BLDM} where the target distribution is factorized such that parallel CCDMs can be used for the constituent binary distributions. Our approach, in contrast, is more general in that the alphabet size per CCDM remains unchanged. MPDM is thus compatible to the aforementioned bit-level distribution matching, and also benefits from low-complexity CCDM implementations, such as those proposed in \cite{Cho2016ECOC_ShapingNonlinearTolerance, Boecherer2017ECOC_StreamingDM,Yoshida2017ECOC_Shaping,Yoshida2018OFC_CCDMAlgo}.

  %% contributions specific: implementation
  In order to simplify the implementation of MPDM, two constraints on the choice of sequences are imposed. Firstly, pairwise partitioning is deployed where each constituent composition (and thus shaped-amplitude sequence) has a complement such that their average has the desired composition. By further requiring the number of sequences of a particular composition to be a power of 2, MPDM with a binary-tree structure is enabled. Hence, MPDM can be implemented by splitting the binary data word into a prefix that selects the composition, and a payload that is mapped onto a symbol sequence in the conventional CCDM fashion. In this paper, we focus on implementation aspects and performance comparisons of distribution matchers. A numerical analysis finds that pairwise tree-based MPDM achieves significantly lower rate loss and thus better AWGN performance than conventional CCDM. To the best of our knowledge, the proposed MPDM is the first distribution matcher that lifts the constant-composition principle for a fixed alphabet size.

\section{Fundamentals of Distribution Matching}

\subsection{Distribution Matching and Notation}\label{ref:ssec:dm}
% We review the principles of CCDM and introduce the notation in the following.

\begin{figure}
\begin{center}
\includegraphics[width=01\columnwidth]{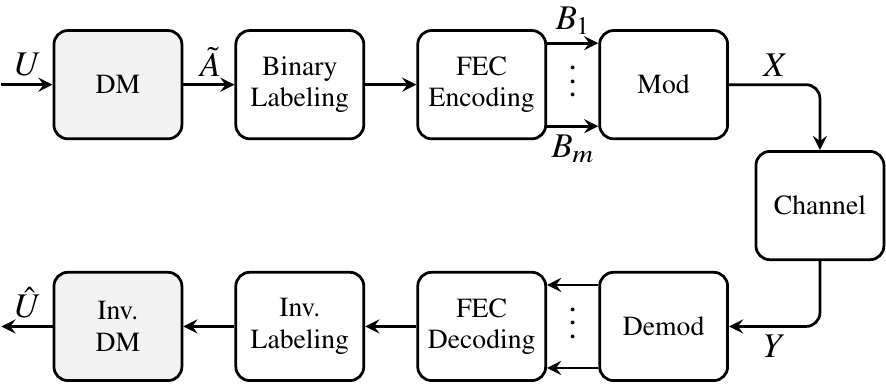}
\end{center}
\vspace{-\baselineskip}
\caption{Diagram of the probabilistic amplitude shaping (PAS) building blocks. This paper studies finite-length distribution matchers (DMs) (gray boxes) that implement an invertible mapping function from the uniform data bits $U$ to the shaped amplitudes $\hat{A}$. On the receiver side, an inverse DM undoes this operation such that $U=\hat{U}$ in the case of error-free FEC output. The PAS logic of combining shaped amplitudes with uniform sign bits is explained in detail in \cite[Sec.~IV]{Boecherer2015TransComm_ProbShaping}.}
\label{fig:block_diagram}
\end{figure}

A DM is an injective mapping from a sequence of length $\Lin$ of uniformly distributed data bits $U$ to $\Lout$ shaped amplitudes. Its integration in the PAS framework is shown in Fig.~\ref{fig:block_diagram}. We consider fixed-length block-wise distribution matching only since variable-length DMs have practical disadvantages such as varying buffer sizes. For finite-length DM, the target distribution $P_A$ must be quantized to $\TypeA$ such that the number of occurrences of each amplitude is integer. Following \cite{Schulte2016TransIT_DistributionMatcher,Boecherer2016TransIT_Quantization}, this quantization is carried out to minimize informational divergence between $P_A$ and $\TypeA$. The resulting amplitudes $\Aquant$ have the probability mass function (PMF) \TypeA, also referred to as type \cite[Sec.~11.1]{CoverThomas2006Book_IT}, \cite[Sec.~II]{Csiszar1998TransIT_Types}, on the alphabet $\SetA=\{a_1,\ldots,a_{\CardA}\}$. %We also loosely use the term composition to mean a type.

Consider a DM output sequence $\Seqx=\{x_1,x_2,\ldots,x_n\}$ of length $\Lout$ where each element $x_j$ with $j\in \{1,\ldots,n\}$ is chosen from \SetA according to $\TypeA$. The number of occurrences $n \left( {a_i} \right)$ of an amplitude $a_i$ in the sequence \Seqx is
\begin{equation}\label{eq:sequence_count}
n \left( {a_i} \right) = \abs{\{\,j: x_j = a_i\,\}}, \quad j\in 1,\ldots,n, \quad i\in 1,\ldots,\CardA, 
\end{equation}
and we have $\sum_{i=1}^{\CardA}n \left( {a_i} \right)=\Lout$. In the following, we write $n_i$ instead of $n \left( {a_i} \right)$ to indicate the number of occurrences of the $i$\textsuperscript{th} amplitude $a_i$. We call the ordered set of occurrences $\Composition = \{n_1,\dots,n_{\CardA}\}$ a \emph{composition}, which has the type $\TypeA$. We say that a sequence has composition \Composition if \eqref{eq:sequence_count} corresponds to \Composition. The set of unique permutations of $\Seqx$ for a given \Composition is referred to as type class \cite[Sec.~11.1]{CoverThomas2006Book_IT}, and its size is the multinomial coefficient \cite[Eq. (11.17)]{CoverThomas2006Book_IT}
\begin{equation}\label{eq:multinom_coeff}
\Npermsfun{\Composition} = \dbinom{n}{n_1,n_2,\ldots,n_{\CardA}} = \frac{n!}{n_1!\, n_2! \cdot \ldots \cdot n_{\CardA}!}.
\end{equation}

\subsection{Constant Composition Distribution Matching (CCDM)}

In the CCDM approach \cite{Schulte2016TransIT_DistributionMatcher}, each of the $2^{\Lin}$ output sequences $\Seqx$ is of type \TypeA, and we denote the single typical CCDM output composition as $\CompositionTypical = \{ n \TypeA(a_1),\ldots,n \TypeA(a_{\CardA})\}$. The constant-composition mapping of a uniform input sequence to a sequence that has \CompositionTypical is denoted as \DMfun{\CompositionTypical} and can, for example, be carried out via arithmetic coding \cite[Sec.~IV]{Schulte2016TransIT_DistributionMatcher}. The number of input bits for CCDM of a particular \CompositionTypical is given by
\begin{equation}\label{eq:Lin_CCDM}
\Lin=\floor{\log_2 \Npermsfun{\CompositionTypical}},
\end{equation}
where $\floor{\cdot}$ denotes rounding down to the closest integer. From \eqref{eq:Lin_CCDM}, we can compute the finite-length rate loss \cite[Eq.~(4)]{Boecherer2017Arxiv_PDM}
\begin{equation}\label{eq:rateloss}
\Rateloss = \Entr{\Aquant} - \frac{\Lin}{\Lout},
\end{equation}
where $\Entr{\cdot}$ denotes entropy in bits. The rate loss vanishes for large \Lout \cite[Eq.~(23)]{Schulte2016TransIT_DistributionMatcher}, which means that an infinite-length CCDM can achieve the target rate without any rate loss. For distribution matching with fixed \Lout and \Composition, it is desirable to make \Lin as large as possible to in order to minimize the rate loss. In the following, we introduce a new class of distribution matcher that has a significantly smaller rate loss than a conventional CCDM.

%%%%%%%%%%%%%%%%%%%%%
%%%%%%Section%%%%%%%%
%%%%%%%%%%%%%%%%%%%%%
\section{Multiset-Partition Distribution Matching}\label{sec:MPDM}
\subsection{Principle}
MPDM is based on the observation that every possible DM output sequence $\Seqx$ is not necessarily of type \TypeA (or equivalently have the composition \Composition) in order to achieve \emph{on average} the target distribution. As the input bits $U$ are uniformly distributed and a DM establishes an injective mapping, it is by the law of large numbers sufficient if the ensemble average of all output sequences has the target composition.\footnote{We note that a rigorous analysis of error exponents for variable-composition DM codewords is an open problem, in particular if there is a one-to-one correspondence between DM and FEC codewords.} Thus, a general MPDM uses those output sequences whose compositions $\Composition_l$ satisfy
\begin{equation}\label{eq:partitioning}
% \Expvalue{P_l} =
% \frac{\sum_l c_l \cdot \Composition_l}{\sum_l c_l} \stackrel{!}{=} \CompositionTypical,
\frac{\sum_l^{\Ncompositions} c_l \cdot \Composition_l}{\sum_l^{\Ncompositions} c_l} \stackrel{!}{=} \CompositionTypical,
% \frac{\sum_l \Npermsfun{\Composition_l} \cdot \Composition_l}{\sum_l \Npermsfun{\Composition_l}} \stackrel{!}{=} \Lout \cdot \TypeA = \CompositionTypical,
\end{equation}
where $l$ indexes the \Ncompositions possible compositions of the MPDM output sequences and
% $\Expvalue{\cdot}$ is the expected value and 
$c_l$ is the number of occurrences of $\Composition_l$ at the MPDM output, with $0 \leq c_l \leq \Npermsfun{\Composition_l}$. The possible compositions $\Composition_l$ can be obtained by exhaustive search, and the choice of $c_l$ depends on the partitioning constraints, see Sec.~\ref{ssec:mpdm:pairwise} for the binary (i.e., pairwise) case. The general partitioning problem \eqref{eq:partitioning} states that the \emph{average} type of all sequences that are the output of a DM must be \TypeA.
The number of distinct compositions is given by
\begin{equation}\label{eq:N_compositions}
\Ncompositions = \dbinom{\Lout + \CardA -1 }{\Lout},
\end{equation}
which can be proven, for example, with the stars-and-bars technique \cite[Sec.~II-5]{Feller1968Book_Probability}.

MPDM can also be viewed in the context of the energy of an $\Lout$-dimensional sphere \cite{Kschischang1993TransIT_Shaping}. In the CCDM approach, all output sequences are on the constant-energy surface of such a sphere in the $\Lout$-dimensional signal space, and, as pointed out in \cite[Sec.~III-C]{GultekinWillems2017PIMRC_EnumerativeShaping}, not all points on that surface are used. In contrast, MPDM combines spheres of different energy levels such that the target distribution $\TypeA$ is achieved at its output.\footnote{This approach is different from indexing energy-bound sequences, as done in \cite{GultekinWillems2017PIMRC_EnumerativeShaping,Schulte2018Arxiv_ShellMapping,GultekinWillems2018ISIT_EnumerativeShaping}. MPDM uses only a subset of sequences, namely those that have a particular composition (and thus energy), which guarantees that the pre-defined target distribution is achieved for the average output sequence.} As will be evaluated in detail in Sec.~\ref{sec:results}, this property gives reduced rate loss compared to CCDM.

\begin{figure}
\begin{center}
\includegraphics[width=01\columnwidth]{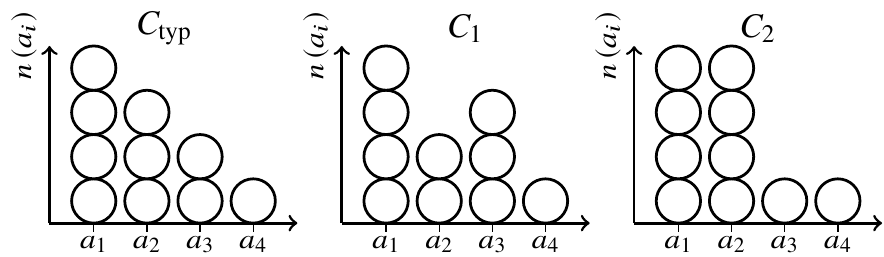}
\end{center}
\vspace{-\baselineskip}
\caption{Illustration of non-constant composition for $\left|\SetA\right|=4$. Combining one sequence that has $\Composition_1$ and one that has $\Composition_2$ gives the target composition \CompositionTypical.}
\label{fig:simple_MPDM_example}
\end{figure}

\begin{example}[Non-Constant Composition]\label{ex:simple_MPDM}
Figure~\ref{fig:simple_MPDM_example} shows a set of figures demonstrating the concept of non-constant composition. We have the typical composition $\CompositionTypical = \{4,3,2,1\}$ for CCDM with $\Lout=10$, which gives the entropy $\Entr{\Aquant}=1.85$~bits. The total number of distinct sequences which have this composition is $\Npermsfun{\CompositionTypical}= 12600$. This determines the rate of the binary distribution matcher to be $\log_2(\floortwo{12600})/10 = 1.3$~bits per symbol. By \eqref{eq:rateloss}, we have a rate loss \Rateloss of $1.85 - 1.3=0.55$~bits for CCDM. By combining one sequence that has $\Composition_1=\{4,2,3,1\}$ and one of $\Composition_2=\{4,4,1,1\}$, the average behavior remains that of the original composition \CompositionTypical. The number of distinct sequences with $\Composition_1$ and $\Composition_2$ are $\Npermsfun{\Composition_1}=12600$ and $\Npermsfun{\Composition_2}=6300$, respectively. If, in addition to \CompositionTypical, these two compositions are used, $c_1=c_2=6300 $~additional sequences can each be generated such that \eqref{eq:partitioning} is fulfilled. Hence, by considering all three compositions in Fig.~\ref{fig:simple_MPDM_example}, we may now use 12600 sequences from \CompositionTypical; 6300 sequences from $\Composition_1$; and 6300 from $\Composition_2$ --- 25200 in total. This increases the rate of the non-constant composition distribution matcher by 0.1~bit/symbol and reduces the rate loss from 0.55~bits to 0.45~bits.
\end{example}

%The problem of designing the MPDM operation can be related to multi-set integer partitioning known from number theory \cite{Wilf2000_LecturesonIntegerPartitions}.

 Suppose the number of input bits $k$ and the DM output length \Lout is fixed. Then, the accumulated composition of all utilized output sequences is $\CompositionAcc = 2^k \cdot \CompositionTypical$. The non-trivial problem is now to find the partitioning of \CompositionAcc into $2^k$ integer subsets while fulfilling \eqref{eq:partitioning} and obeying two constraints in the subset selection: the sum of the integer elements in each subset must be equal to \Lout in order to have a fixed-length DM, and each subset cannot occur more often than their multinomial coefficient \Npermsfun{\Composition} (see \eqref{eq:multinom_coeff}) such that an injective mapping function is established.

\begin{example}[General MPDM]\label{ex:partitioning}
  Consider a DM with $k=17$, $\Lout=10$, and $\TypeA=[0.4, 0.3, 0.2, 0.1]$. We have $\CompositionTypical = \{4,3,2,1\}$, and the accumulated composition is $\CompositionAcc = \{524288,393216,262144,131072\}$. A general MPDM seeks to find those $2^{17}$ out of the $2^{20}$ possible DM output sequences whose number of occurrences of each amplitude gives \CompositionAcc, i.e., that fulfills \eqref{eq:partitioning}. By \eqref{eq:N_compositions}, there are $\Ncompositions=286$~compositions for these sequences.
  % Note that a brute-force solution is clearly infeasible as even for this short-DM case, there are $\binom{2^{20}}{2^{17}} \approx 5.3\times10^{171574}$ ways of choosing the sequences. 
  The problem is equivalent to finding the non-unique integer sets (each of which corresponds to a particular \Composition) that sum up to \CompositionAcc, given the constraints that the sum over each set must be $\Lout=10$ and that each set occurs at most $\Npermsfun{\Composition}$ times.
\end{example}

Many partitioning problems are known to be NP-complete \cite[Sec.~3.1.5]{Garey1979Book_NPComplete}, yet algorithms giving approximate solutions with reasonable complexity are known for special cases \cite[Sec.~4.2]{Garey1979Book_NPComplete}. While the exact complexity of the considered constrained multiset partitioning problem is unknown to us, it also irrelevant since finding a solution does not lead to a constructive algorithm for MPDM design. In other words, even if a solution to the partitioning problem could be found, it would remain a challenging task to establish an implementable mapping function between DM input and output sequences, in particular if the DM dimensions prohibit the use of a lookup table. By imposing some structure onto the partitioning, a construction of a MPDM device is made feasible at the expense of a slightly increased rate loss, as we will show next.

\subsection{Pairwise MPDM}\label{ssec:mpdm:pairwise}

To facilitate the implementation of MPDM, we simplify the general partitioning problem \eqref{eq:partitioning} by considering pairwise typical sequences only. Note that other structured partitioning schemes, for instance into triples or quadruples, are also possible.\footnote{With MPDM, \TypeA can also be requantized such that the divergence between the target PMF and \TypeA is reduced. Potential benefits of this approach remain for future work.} In this pairwise case, we require that for every composition $\Composition_l$, a complementary composition $\CompositionComplement_l$ must exist such that
\begin{equation}\label{eq:pair_plus_complement}
\Composition_l + \CompositionComplement_l = 2 \cdot \CompositionTypical.
\end{equation}
The unique pairs can be found in a relatively straightforward fashion by exhaustive search, i.e., by cycling through all possible compositions (of which there are \Ncompositions as per \eqref{eq:N_compositions}) and discarding those that, when added to their complement as per \eqref{eq:pair_plus_complement} do not give \CompositionTypical. The number of valid pairs, denoted as \CompositionPair, that fulfill the relation \eqref{eq:pair_plus_complement} can be computed by modifying \eqref{eq:N_compositions} with the inclusion-exclusion method \cite[Theorem~4.2]{Allenby2011Book_HowToCount}, taking into account that certain compositions can never occur in a constrained setting such as the considered pairwise MPDM. The idea of inclusion-exclusion is to start with the unconstrained number of compositions \Ncompositions as per \eqref{eq:N_compositions} and remove from it those compositions that, on a single amplitude basis, never lead to a the desired distribution. Next, invalid compositions for all pairwise combinations of two amplitudes are included as they have been excluded twice in the previous step. Triple-wise combinations must be excluded again, and so forth. This alternation between inclusion and exclusion is repeated \CardA times.

\begin{example}[Inclusion-Exclusion]\label{ex:inclusion-exclusion}
  Consider the case of Example~\ref{ex:simple_MPDM} where $\CompositionTypical = \{4,3,2,1\}$ and $\Lout=10$. According to \eqref{eq:N_compositions}, we have $\Ncompositions = 286$. For the first amplitude $a_1$, compositions with more than 8 occurrences of $a_1$ cannot be combined in a pairwise manner while fulfilling \eqref{eq:pair_plus_complement}. Thus, the four compositions with either 9 or 10 occurrences of $a_1$ are excluded. This is repeated for $a_2$, $a_3$, and $a_4$, resulting in a respective reduction of valid compositions by 20, 56, and 120, which gives an interim composition count of 86. However, some compositions have been excluded twice (such as $\Composition=\{0,7,0,3\}$ violating both the $a_2$ and the $a_4$ constraint), and thus have to be included once again. In total, there is 11 such excessive exclusions, giving a final composition count of 97 for pairwise partitioning. This corresponds to 49 distinguishable pairs out of which one is the degenerate CCDM ``pair'' $\{\CompositionTypical,\CompositionTypical\}$. Note that triple- and quadruple-wise combinations of amplitudes do not have to be considered in this simple example.
\end{example}
%, taking into account that certain compositions can never occur in a constrained setting, such as the considered pairwise MPDM.
For pairwise partitioning, the MPDM output sequences that have $\Composition_l$ or $\CompositionComplement_l$ should occur in an equiprobable manner, which implies that the total number of permutations for a pair is governed by the constituent composition that has fewer permutations. We denote the permutation count of a pair \CompositionPair as
\begin{equation}\label{eq:count_pairwise}
\hspace{-1pt} \Npermsfun{\CompositionPair} = \left\{ 
\begin{array}{lr}
2 \cdot \min\left(\Npermsfun{\Composition^{\phantom{\prime}}_l}, \Npermsfun{\CompositionComplement_l}\right),& \Composition_l \neq \CompositionComplement_l, \\
\Npermsfun{\CompositionTypical},& \Composition_l = \CompositionComplement_l,
\end{array}
\right.
\end{equation}
where the first case corresponds to non-degenerate pairs and the latter case is the degenerate CCDM ``pair''.
For a pairwise MPDM with \Npairs distinguishable pairs \CompositionPair that satisfy \eqref{eq:pair_plus_complement}, the total number of permutations is
\begin{equation}\label{eq:perms_NMPDM}
\Nperms = \sum_{l=1}^{\Npairs} \Npermsfun{\CompositionPair}.
% \sum_{l=1}^{\Npairs} \underbrace{2 \cdot \min\left(\Npermsfun{\Composition_l}, \Npermsfun{\CompositionComplement_l}\right)}_{\text{non-degenerate pairs:} \Composition_l \neq \CompositionComplement_l~\forall l} + \underbrace{\Npermsfun{\CompositionTypical}}_{\text{CCDM}}.
\end{equation}
Note that \CompositionPair is invariant to permutations of the compositions and switching the two compositions (i.e., $\{\CompositionComplement_l, \Composition_l\}$ instead of \CompositionPair) does not give a new unique pair. The rate loss improvement of MPDM over CCDM (see Sec.~\ref{sec:results}) is the result of including non-degenerate pairs in \eqref{eq:perms_NMPDM} in addition to the typical CCDM composition.\footnote{It is only in the trivial case where $\CompositionTypical$ has only one non-zero element that the number of permutations of CCDM and MPDM is identical.} For any DM with binary input, \eqref{eq:perms_NMPDM} is rounded down to the nearest power of 2, i.e., we have
\begin{equation}\label{eq:perms_NMPDM_floor2}
2^\Lin = \floortwo{\Nperms}.
% \sum_{l=1}^{\Npairs} \underbrace{2 \cdot \min\left(\Npermsfun{\Composition_l}, \Npermsfun{\CompositionComplement_l}\right)}_{\text{non-degenerate pairs:} \Composition_l \neq \CompositionComplement_l~\forall l} + \underbrace{\Npermsfun{\CompositionTypical}}_{\text{CCDM}}.
\end{equation}
The same requirement is made for binary CCDM in \eqref{eq:Lin_CCDM}.
\begin{example}[Pairwise MPDM]\label{ex:MPDM}
  Consider $P_A = [0.4415, 0.3209, 0.1654, 0.0722]$ (taken from \cite[Example~A]{Schulte2016TransIT_DistributionMatcher}) and \Lout = 10. We use \cite[Algorithm~2]{Boecherer2016TransIT_Quantization} to quantize $P_A$ to $\TypeA=[0.4, 0.3, 0.2, 0.1]$, which has $\Entr{\Aquant}=1.85$~bits and $\CompositionTypical = \{4,3,2,1\}$, see Example~\ref{ex:simple_MPDM}. For pairwise MPDM, there are $\Npairs=49$~pairs (including the degenerate one) that fulfill \eqref{eq:pair_plus_complement}, see Example~\ref{ex:inclusion-exclusion}. The new total permutation count is 164214, which increases the number of input bits to $\Lin=17$ and thus reduces the rate loss to 0.15~bits.
\end{example}

Although considering pairwise-typical compositions greatly simplifies the search for valid partitionings, the implementation of such a pairwise MPDM is not straightforward if a large lookup table is not to be used. In the following, we impose another constraint, again at the expense of transmission rate, that enables implementation of MPDM with reasonable complexity.

\subsection{Implementation with a Binary Tree Structure}
\begin{figure}
\begin{center}
\includegraphics[width=01\columnwidth]{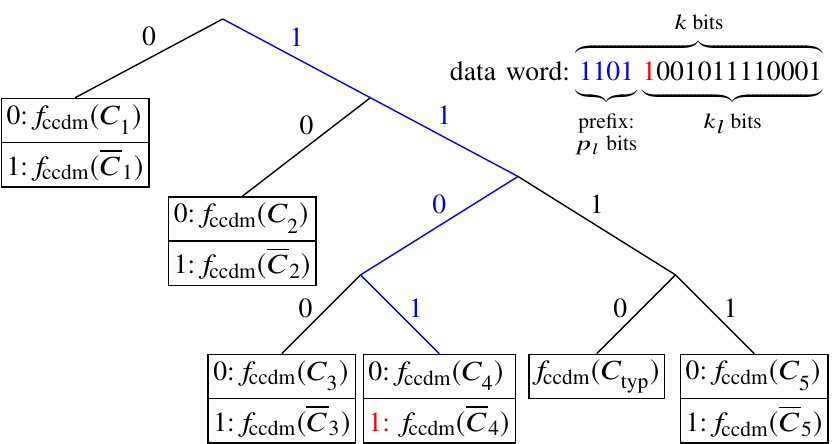}
\end{center}
\caption{Illustration of a tree-structured pairwise MPDM with six pairs, one of which corresponding to the degenerate composition \CompositionTypical. The mapping operation for a 17-bit input sequence (top right) to the composition $\CompositionComplement_4$ is exemplified. A prefix of $p_l=4$ bits length (blue) chooses the pair. The next bit (red) selects the composition within the pair, which is omitted for \CompositionTypical. Mapping the 12-bit payload (black) onto a sequence of shaped amplitudes that has $\CompositionComplement_4$ can be carried out with a conventional CCDM algorithm, such as arithmetic coding.}
\label{fig:tree_structure}
\end{figure}
% The implementation of a pairwise MPDM in a tree-like structure is outlined in the following.
Any DM with binary input uses $\floortwo{\Nperms}$ out of \Nperms output sequences. In order to design an implementable pairwise MPDM, we require, in addition to \eqref{eq:pair_plus_complement}, that the number of permutations $\Npermsfun{\CompositionPair}$ of each pair \CompositionPair must be a power of 2. A pair \CompositionPair thus represents an integer $k_l$ bits, and \eqref{eq:count_pairwise} becomes
\begin{equation}\label{eq:tree_req}
2^{k_l} = \left\{ 
\begin{array}{lr}
2 \cdot \min\left(\floortwo{\Npermsfun{\Composition^{\phantom{\prime}}_l}}, \floortwo{\Npermsfun{\CompositionComplement_l}}\right),& \, \Composition_l \neq \CompositionComplement_l, \\
\floortwo{\Npermsfun{\CompositionTypical}},& \, \Composition_l = \CompositionComplement_l.
\end{array}
\right.
\end{equation}
% \begin{equation}\label{eq:tree_req}
% 2^{k_l} = 2 \cdot \min\left(\floortwo{\Npermsfun{\Composition_l}}, \floortwo{\Npermsfun{\CompositionComplement_l}}\right)
% \end{equation}
% for the non-degenerate pairs. For the typical CCDM composition \CompositionTypical, we write 
% \begin{equation}\label{eq:tree_req_typ}
% 2^{k_\text{typ}} = \floortwo{\Npermsfun{\CompositionTypical}}.
% \end{equation}
The total number of permutations with this power-of-2 constraint is thus
\begin{equation}\label{eq:tree_Nperms}
% \Nperms =
2^\Lin = \sum_{l=1}^{\NpairsReduced} 2^{k_l},
\end{equation}
where \NpairsReduced out of the initially available \Npairs pairs are selected as to maximize \Nperms (while keeping it a power of 2 to have a binary DM) and thus, to maximize \Lin. The selection of pairs can be done by sorting \CompositionPair according to $k_l$ in descending order and including only the first \NpairsReduced pairs in that ranked list until \Lin is integer. We note that the constraint \eqref{eq:tree_req} can lead to fewer permutations than for unconstrained pairwise MPDM (see \eqref{eq:perms_NMPDM}) and thus an increased rate loss.

\begin{example}[Pairwise MPDM with Binary-tree Structure]\label{ex:MPDM_tree}
Consider the case of Example \ref{ex:MPDM}. With \eqref{eq:tree_req} and \eqref{eq:tree_Nperms}, the total number of permutations is computed to be 122688, which gives $k=16$ input bits and $\Rateloss=0.25$~bits, see the marker in Fig.~\ref{fig:ccdm_comparison}. The number of pairs that is necessary to address the 16~bits is $\NpairsReduced=9$ out of the initial $\Npairs = 49$. Note that \CompositionTypical is not included in these 9 pairs as they already maximize the integer-valued \Lin.
\end{example}

%% tree general
We outline in the following how a binary tree with variable-length prefix $p_l$ can be constructed such that individual compositions can be addressed with a constant \Lin. Dividing \eqref{eq:tree_Nperms} by $2^\Lin$ gives
\begin{equation}\label{eq:tree_Nperms_Kraft}
\sum_{l=1}^{\NpairsReduced} 2^{-(k-k_l)} = 1,
\end{equation}
which is the Kraft inequality \cite[Theorem~5.2.1]{CoverThomas2006Book_IT} for a binary alphabet and fulfilled with equality. Thus, the lengths of the prefixes and of the bits to be mapped always sum up to $\Lin$:
\begin{equation}\label{eq:tree_prefix}
p_l + k_l = k,~~ l=\{1,\ldots,\NpairsReduced\}.
\end{equation}
By \eqref{eq:tree_Nperms}--\eqref{eq:tree_prefix}, we are effectively enforcing a selection of the compositions that follows a dyadic distribution and use a Huffmann code for determining the prefix \cite[Sec.~5.6]{CoverThomas2006Book_IT}. The probability of the composition pair $\CompositionPair$ being selected is $2^{-p_l}$ such that $p_l$ bits of prefix can address it without loss.

The power-of-2 constraint of \eqref{eq:tree_Nperms} thus enables the implementation of MPDM in a binary-tree structure as follows. The \NpairsReduced different pairs are sorted by their $k_l$ in ascending order, and the two compositions within a pair are labeled 0 and 1, respectively. Note that this single-bit label is omitted in the special case of the CCDM composition \CompositionTypical. The two pairs with the smallest $k_l$ (i.e., the least permutations) form a branch, with one element labeled 1 and the other 0. If more than one branch remains, i.e., if there are pairs that have not been used in the tree yet, the merging and labeling process is repeated. When only a single branch remains, the prefix tree is completed. This standard source-coding technique gives an optimal labeling for the binary tree.

%% tree input
Once this binary tree is generated, the mapping from $\Lin$-bit uniform data word to shaped amplitude sequence is done by splitting the MPDM input sequence into three parts. The first $p_l=\Lin-k_l$ bits are the prefix that identifies the pair. The next bit chooses the composition within that pair. For the mapping of the final $k_l-1$ bits onto the shaped symbols according to the selected composition, conventional CCDM based on arithmetic coding can be employed \cite[Sec.~IV]{Schulte2016TransIT_DistributionMatcher}. This tree-structure design is illustrated in Fig.~\ref{fig:tree_structure}.

% Each pair can be addressed with $k_l$ bits, and the remaining $k-k_l$ bits are represented by which pair is chosen. 

At the receiver, inverse MPDM of a shaped sequence $\Seqx$ must be performed in order to recover the initially transmitted data word. Note that MPDM is designed as an invertible function and hence will not introduce any errors if the input, i.e., the FEC decoder output, is error-free. First, the composition of $\Seqx$ is determined by, e.g., a simple histogram operation, from which the binary prefix of length $k-k_l+1$ can be looked up. In order to obtain the remaining $k_l-1$ payload bits, an inverse CCDM algorithm using arithmetic coding can be employed. This recovers the transmitted bit sequence. Note that the entire MPDM codeword must be detected before inverse MPDM mapping can begin because the composition determines the prefix. For potential real-time processing, demapping therefore must be parallel on a per-codeword basis, i.e., processing the next codeword must begin before the current one finishes.

The key steps for constructing a MPDM  are summarized in Algorithm~\ref{alg:MPDM}. The resulting MPDM of rate $\frac{\Lin}{\Lout}$ will have pairwise partitioning and use a binary-tree structure for selecting the component compositions.

\begin{algorithm}
  \caption{Construction of a Pairwise Binary-Tree MPDM}\label{alg:MPDM}
  \begin{algorithmic}[1]
  \Require \Lout, $P_A$ \Comment{MPDM output length, target distribution}
    \State Determine typical composition \CompositionTypical \Comment{See Sec.~\ref{ref:ssec:dm}}
    \State Find all composition pairs \CompositionPair with $l=\{1,\ldots,\Npairs\}$ %\Comment{See Sec.~\ref{ssec:mpdm:pairwise}}
    \State Restrict usage per pair to largest power of two that is not greater than $2\cdot \min\left(\Npermsfun{\Composition^{\phantom{\prime}}_l}, \Npermsfun{\CompositionComplement_l}\right)$ \Comment{See \eqref{eq:tree_req}} \label{alg:MPDM:p2Per}
    \State Sort the pairs in descending order by the usage count of step \ref{alg:MPDM:p2Per}
    \State Choose the first $N^\dagger_\text{pairs}$ such that the total number of MPDM permutations is $2^k$ where $k$ is as large as possible\label{alg:MPDM:select_p2_pairs}
    \State Each pair \CompositionPair chosen in step \ref{alg:MPDM:select_p2_pairs} is assigned a prefix of length $k-k_l$ bits 
    \State A one-bit prefix determines the composition within a pair   
  \end{algorithmic}
\end{algorithm}
% }

In the next section, we compare pairwise MPDM with a binary-tree-structured implementation (which we simply refer to as MPDM) to conventional CCDM.

%%%%%%%%%%%%%%%%%%%%%
%%%%%%Section%%%%%%%%
%%%%%%%%%%%%%%%%%%%%%
\section{Numerical Results}\label{sec:results}
This section numerically studies the rate loss and AWGN performance of pairwise MPDM with the tree-structure design outlined in Sec.~\ref{sec:MPDM}. Rate loss is computed with \eqref{eq:rateloss}, where the input length \Lin is computed from \eqref{eq:Lin_CCDM} for CCDM and from \eqref{eq:tree_Nperms} for MPDM. For the AWGN channel results, we consider quadrature amplitude modulation (QAM) channel input as the concatenation of two one-dimensional amplitude-shift keying (ASK) symbols. The figure of merit is the achievable information rate (AIR) for bit-metric decoding reduced by the DM rate loss,
\begin{equation}\label{eq:Rbmd}
\AIRFiniteDM = \left[ \Entr{\mathbf{B}} - \sum_{i=1}^{m} \Entr{B_i|Y} \right]- \Rateloss,
\end{equation}
% where $\mathbf{B} = [ B_1,B_2,\ldots,B_{\log_2M} ]$ is the vector of shaped bits (which, taken in groups of 1D-amplitudes, follows $\TypeA$) and $Y$ is the complex AWGN output. The factor 2 is included because \eqref{eq:rateloss} is defined in bits per 1D amplitude.
where $\mathbf{B}=[B_1,\ldots,B_M]$ describes the binary input into the modulator and $Y$ the symbolwise channel output, see Fig.~\ref{fig:block_diagram}. The derivation of \eqref{eq:Rbmd} is given in the Appendix. A quantized version of the optimal Maxwell-Boltzmann distribution \cite[Sec.~IV]{Kschischang1993TransIT_Shaping} for each SNR is used as \TypeA. We focus on 64QAM for the AWGN rate analysis, emphasizing that MPDM is feasible with any modulation format that is compatible with PAS.

\begin{figure}
\begin{center}
% \inputtikz{Rate_CCDM_PDM}
\includegraphics[width=01\columnwidth]{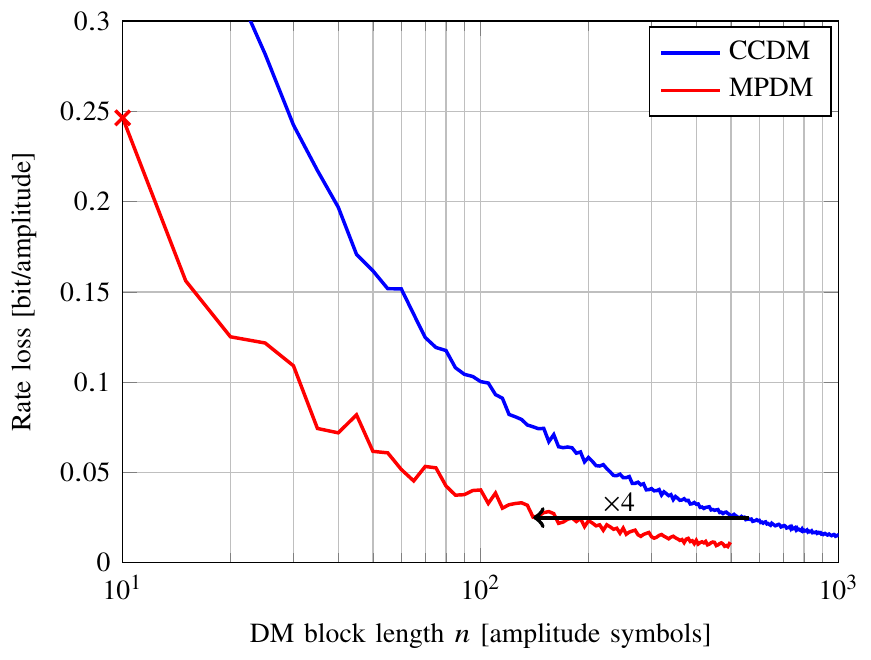}
\end{center}
\caption{Rate loss over block length for conventional CCDM and pairwise MPDM implemented with a tree structure. The target PMF is $P_A = [0.0722, 0.1654, 0.3209, 0.4415]$ from \cite[Example A]{Schulte2016TransIT_DistributionMatcher}. The marker for MPDM at $\Lout=10$ refers to Example~\ref{ex:MPDM_tree}.} 
\label{fig:ccdm_comparison}
\end{figure}

Fig.~\ref{fig:ccdm_comparison} shows rate loss over block length for the target PMF of \cite[Example A]{Schulte2016TransIT_DistributionMatcher}. We observe that the pairwise MPDM achieves a smaller rate loss compared to CCDM for all block lengths. For a rate loss of 0.025~bits per amplitude symbol, MPDM can operate with approximately $\Lout = 140$ symbols, whereas a conventional CCDM requires a fourfold increase in length. Note that the jagged shape of CCDM and MPDM rate loss is due to flooring operations in \eqref{eq:Lin_CCDM} and \eqref{eq:perms_NMPDM_floor2}, respectively.

% \subsection{Information Rates for AWGN}
\begin{figure}
\begin{center}
\includegraphics[width=01\columnwidth]{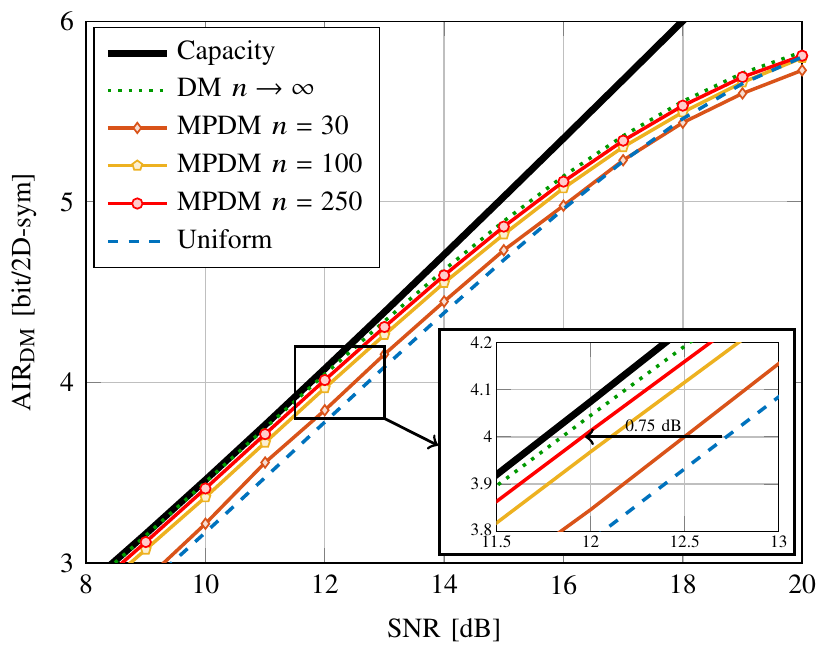}
\end{center}
\caption{AIR in bit/2D-sym over SNR in dB for bit-metric decoding and 64QAM. The AWGN capacity $\log_2(1+\text{SNR})$ is shown as reference. The inset zooms into the region around $\AIRFiniteDM = 4$~bit/2D-sym where MPDM of length $n=250$ is 0.75~dB more power-efficient than uniform 64QAM.}
\label{fig:gmi_snr}
\end{figure}

\begin{figure}
\begin{center}
\includegraphics[width=01\columnwidth]{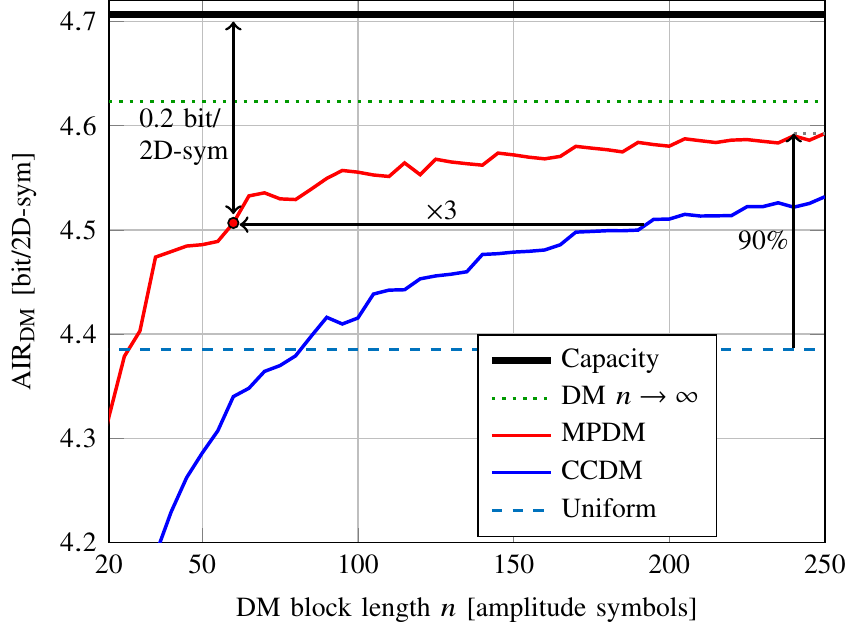}
\end{center}
\caption{AIR in bit/2D-sym over the block length \Lout for 64QAM at 14~dB SNR. MPDM with $\Lout=60$ (marker) operates within 0.2~bit/2D-sym of capacity, whereas a conventional CCDM requires three times the length. At $\Lout=250$, the MPDM achieves 90\% of the maximum available shaping gain at this SNR, which is given by an infinite-length DM.}
\label{fig:gmi_blocklength}
\end{figure}

In Fig.~\ref{fig:gmi_snr}, \AIRFiniteDM in bits per 2D-symbol (bit/2D-sym) is shown over SNR in dB for 64QAM. In addition to the AIRs for MPDM of short ($\Lout=30$), medium ($\Lout=100$), and large size ($\Lout=250$), the AWGN capacity (solid black), an infinite-length DM without rate loss (dotted) and uniform 64QAM (dashed) are included as references. We observe that MPDM with length as small as $\Lout=30$ has larger AIR than uniform 64QAM over the relevant SNR range. By increasing $\Lout$ to 250~symbols, the MPDM achieves performance within 0.2~dB of the Shannon bound at an AIR of 4~bit/2D-sym.

In Fig.~\ref{fig:gmi_blocklength}, a comparison of MPDM and CCDM as a function of the block length \Lout in 1D amplitude symbols is shown for a fixed SNR of 14 dB. At this SNR level and for 64QAM, an infinite-length CCDM without any rate loss operates within approximately 0.1~bit/2D-sym of the AWGN capacity. For MPDM lengths above 30 symbols, we note that performance is better than that of uniform 64QAM, while CCDM requires at least 80 symbols to overcome the rate loss. MPDM with $\Lout=60$ is able to operate within 0.1~bit/2D-sym of the infinite-length limit and thus within 0.2~bit/2D-sym of capacity. We further note that for $\Lout=60$, MPDM achieves half of the available shaping gain of 0.24~bit/2D-sym. By increasing the MPDM length to $\Lout=250$, 90\% of the shaping gain are attainable.

\begin{figure}
\begin{center}
\includegraphics[width=01\columnwidth]{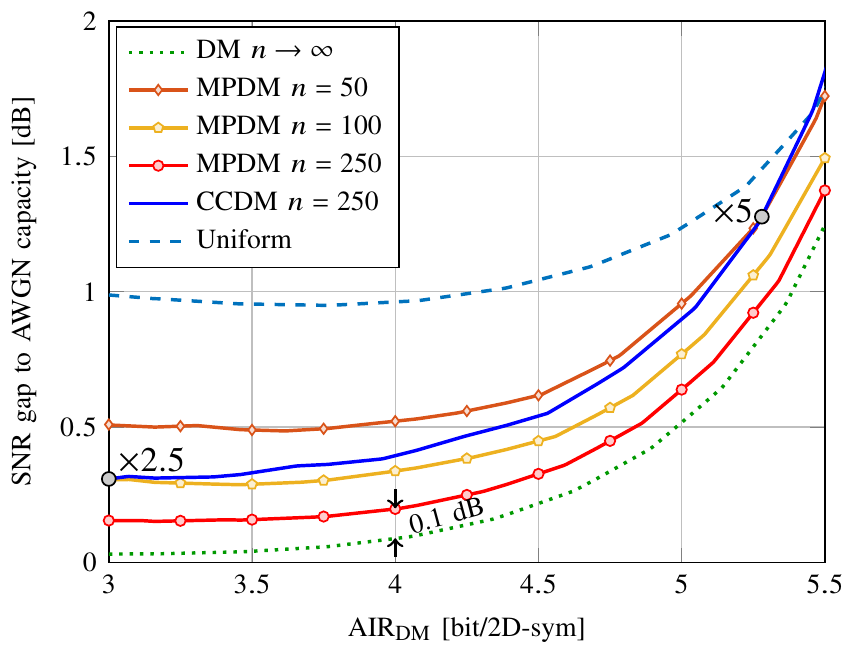}
\end{center}
\caption{SNR gap to AWGN capacity in dB over AIR. The gray markers show that the block length reduction of MPDM compared to CCDM is between a factor of 2.5 and 5.}
\label{fig:SNRgap_AIR}
\end{figure}

Fig.~\ref{fig:SNRgap_AIR} shows the SNR gap to capacity in dB over \AIRFiniteDM. For all considered rates, a length-250 MPDM operates within approximately 0.1~dB of its asymptotic limit. When comparing CCDM with $\Lout = 250$ to MPDM of various lengths, we observe that the MPDM length reduction is approximately a factor of 2.5 at low AIRs (left gray marker). At $\AIRFiniteDM=3$~bit/2D-sym, MPDM of length 100 has an SNR gap to capacity of 0.3~dB. When increasing \AIRFiniteDM, CCDM approaches (and eventually crosses) the MPDM curve of $\Lout=50$, which corresponds to a fivefold length reduction (right gray marker). For large AIRs beyond 5.5~bit/2D-sym (not shown) where the QAM PMF is close to uniform, the length reduction is up to a factor of 10, indicating that the MPDM benefit depends on how strongly the quantized PMF is shaped. For a heavily shaped distribution, pairwise MPDM only gives few additional permutations over a conventional CCDM. As we have seen from the results in this section, this number and thus the potential input sequence \Lin increases drastically when the PMF is closer to a uniform one, leading to superior performance of MPDM over CCDM.

\section{Conclusion}
We have proposed a novel distribution matching scheme, referred to as multiset-partition distribution matching (MPDM), which generalizes conventional CCDM by lifting the constant-composition property of the shaped distribution matcher output sequences. By including sequences with many different compositions which have the desired composition on average, MPDM achieves a lower rate loss than CCDM for a fixed block length in all relevant cases of distribution matching. By imposing constraints on the choice of partitionings and their number of occurrences, a computationally efficient, constructive algorithm for distribution matching and dematching is devised. MPDM is numerically found to allow the block length to be reduced by a factor of 4 for the same rate loss as CCDM. AWGN simulations with 64QAM demonstrate that this reduction depends on the SNR (i.e., the target distribution) and amounts to a factor of 2.5 to 5 for a fixed gap to AWGN capacity.

\section*{Acknowledgments}
The authors would like to thank the anonymous reviewers for their comments which greatly helped to improve the presented paper.

\appendix[AIR Evaluation for Finite-length DM]\label{app:gmi_rateloss}

\begin{figure}
\begin{center}
\includegraphics[width=01\columnwidth]{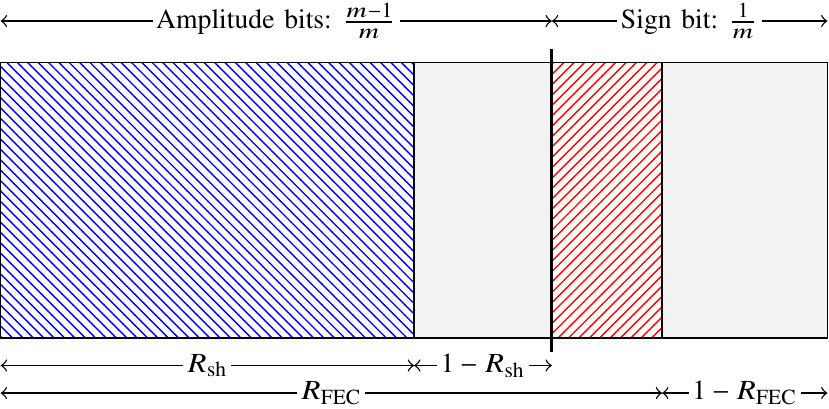}
\end{center}
\caption{Composition of an ASK symbol with shaping according to the PAS scheme. The fixed PAS boundary (thick vertical line) between $m-1$ shaped amplitude and $1$ uniform sign bit is shown for 8ASK ($m=3$). The striped areas show the information in the amplitude bits (blue) and in the sign bit (red). The gray areas represent the redundancy of the shaping code and the FEC code, respectively.}
\label{fig:PAS_Codeword_Breakdown}
\end{figure}

The following derivation shows how an achievable information rate (AIR) for a finite-length DM can be computed from the conventionally estimated AIR (assuming infinite-length DM) and the DM rate loss. Both DM and FEC are considered as codes with rates \Rshaping and \RFEC, respectively. We then evaluate the information content per ASK symbol for an infinite-length and finite-length DM.

In the PAS scheme, $m-1$ shaped amplitude bits of each ASK symbol are combined with $1$ uniform sign bit. The source of these sign bits can be the uniform data that is to be transmitted or the parity bits of the FEC code. A schematic illustration of such an average ASK-symbol composition is given in Fig.~\eqref{fig:PAS_Codeword_Breakdown}. The combined striped areas represent the overall amount of information contained in each symbol. This information content \InfoRcombined in bits per ASK symbol is
\begin{equation}\label{eq:PAS_RateCombined}
\InfoRcombined = \left[\frac{m-1}{m} \cdot \Rshaping + \left( \RFEC - \frac{m-1}{m}\right)\right]\cdot m,
\end{equation}
where the first term inside the brackets corresponds to the information contained in the shaped amplitudes (blue striped area in Fig.~\eqref{fig:PAS_Codeword_Breakdown}) and the second term is the information within the sign bits (red striped area), sometimes denoted $\gamma$ \cite[Sec.~IV-D]{Boecherer2015TransComm_ProbShaping}. For PAS, \RFEC must be at least $\frac{m-1}{m}$ \cite[Sec.~IV-B]{Boecherer2015TransComm_ProbShaping}. The shaping rate \Rshaping in \eqref{eq:PAS_RateCombined} is defined as
\begin{equation}\label{eq:rshaping}
\Rshaping = \left\{ \begin{array}{lr}
\frac{\frac{k}{n}}{m-1} & \text{for finite-length DM},\\[5pt]
\frac{\Entr{\Aquant}}{m-1} & \text{for infinite-length DM},
\end{array}
\right.
\end{equation}
respectively, and thus describes the ratio of information contained in the shaped amplitude bits. With the definitions \eqref{eq:PAS_RateCombined} and \eqref{eq:rshaping}, we can state a performance measure of a finite-length DM. We define the efficiency $\eta$ of a finite-length DM as the ratio of \InfoRcombined for finite-length and infinite-length DM, i.e.,
\begin{align}
\Effshaping ={}& \frac{\frac{m-1}{m} \cdot \frac{\frac{k}{n}}{m-1} + \left( \RFEC - \frac{m-1}{m}\right)}
{\frac{m-1}{m} \cdot \frac{\Entr{\Aquant}}{m-1} + \left( \RFEC - \frac{m-1}{m}\right)} \\
% ={}& \frac{\frac{k}{n} + m\cdot\RFEC - m + 1 } {\Entr{\Aquant} + m\cdot\RFEC - m + 1 } \\
={}& \frac{\frac{k}{n} + 1 + m \cdot \left( \RFEC - 1 \right)} {\Entr{\Aquant} + 1 + m \cdot \left( \RFEC - 1 \right)}.
\label{eq:shaping_eff}
\end{align}
In the following, we use this shaping efficiency to compute AIRs for finite-length DM.

Consider an AIR in bits per symbol that is computed without including any DM rate loss. A highly relevant AIR for PAS and binary FEC is the bit-metric decoding (BMD) rate \Rbmd defined as \cite[Eq. (63)]{Boecherer2015TransComm_ProbShaping}
\begin{equation}\label{eq:Rbmd_inflength}
\Rbmd = \left[ \Entr{\mathbf{B}} - \sum_{i=1}^{m} \Entr{B_i|Y} \right],
\end{equation}
where $\mathbf{B}=\left[B_1,B_2,\ldots,B_m\right]$ is the binary input vector and $Y$ the channel output (see Fig.~\ref{fig:block_diagram}). The computation of \Rbmd can, for example, be carried out via numerical integration if the channel law is known, or in Monte Carlo simulations and by mismatched decoding for an unknown channel \cite[Sec.~VI]{Arnold2006TransIT_AchievableRates}. We use numerical integration for the AWGN results in Sec.~\ref{sec:results}. Note that \Rbmd according to \eqref{eq:Rbmd_inflength} is achievable only for a DM without rate loss. The BMD rate for PAS and a finite-length DM, referred to as \AIRFiniteDM, is given by
\begin{equation}\label{eq:AIR_multiplic}
\AIRFiniteDM = \Effshaping \cdot \Rbmd.
\end{equation}
This means that by using a finite-length DM, the information content of each symbol is reduced by $\eta$. Note that \eqref{eq:AIR_multiplic} also holds for AIRs other than \Rbmd, such as mutual information. To simplify \eqref{eq:AIR_multiplic}, we consider PAS with capacity-achieving codes that operate at their thresholds, in which case we have
\begin{equation}
\Rbmd \equiv \InfoRcombined.
\end{equation}
Solving \eqref{eq:PAS_RateCombined} for the FEC rate then gives
\begin{equation}\label{eq:FEC_asympt}
\RFEC = \frac{\Rbmd}{m} + \left(1-\frac{\Entr{\Aquant}}{m-1} \right)\cdot \frac{m-1}{m}.
\end{equation}
By inserting \eqref{eq:FEC_asympt} into \eqref{eq:shaping_eff} we get 
\begin{equation}\label{eq:shaping_eff_FEC_asympt}
\Effshaping = 1 + \frac{\frac{k}{n} - \Entr{\Aquant}}{\Rbmd},
\end{equation}
which, inserted into \eqref{eq:AIR_multiplic}, finally gives a simplified expression for the BMD rate of a finite-length DM,
\begin{align}
\AIRFiniteDM ={}& \Rbmd + \frac{\Lin}{\Lout} - \Entr{\Aquant} \\
={}& \Rbmd - \Rateloss, \label{eq:AIR_rateloss_absolute}
\end{align}
where \Rateloss was introduced in \eqref{eq:rateloss}. We use \eqref{eq:AIR_rateloss_absolute} in Sec.~\ref{sec:results} to compare the AIRs of DMs that have different rate losses, in particular of a conventional CCDM and the proposed MPDM.

%%%%%%%%%%%%%%%%%%%%%%%%%%%%%%%%%%%%%%%%%%%
%%              References               %%
%%%%%%%%%%%%%%%%%%%%%%%%%%%%%%%%%%%%%%%%%%%
\bibliographystyle{IEEEtran}
\bibliography{myJournalsFull,references,myPublications}

\end{document}